\documentclass[aps,pra,twocolumn,showpacs,superscriptaddress]{revtex4-2}

\usepackage{amsfonts,mathrsfs,amsmath,amsthm,amssymb}
\usepackage{txfonts}
\usepackage{bm,times}
\usepackage{graphicx,epsfig}
\usepackage[colorlinks=true,breaklinks=true,linkcolor=blue,citecolor=blue,urlcolor=blue]{hyperref}

\def \tr{{\rm{Tr}}}

\newcommand{\be}{\begin{eqnarray}}
\newcommand{\ee}{\end{eqnarray}}

\makeatletter
    
    \newcommand{\Rmnum}[1]{\expandafter\@slowromancap\romannumeral #1@}
\makeatother

\begin{document}
\author{Guangming Jiang}
\affiliation{College of Physical Science and Technology, Sichuan University, Chengdu 610064, China}

\author{Xiaohua Wu}
\email{wxhscu@scu.edu.cn}
\affiliation{College of Physical Science and Technology, Sichuan University, Chengdu 610064, China}

\author{Tao Zhou}
\email{taozhou@swjtu.edu.cn}
\affiliation{Quantum Optoelectronics Laboratory, School of Physical Science and Technology, Southwest Jiaotong University, Chengdu 610031, China}
\affiliation{Department of Applied Physics, School of Physical Science and Technology, Southwest Jiaotong University, Chengdu 611756, China}

\date{\today}

\title{Quantum steering in a star network}

\begin{abstract}
In this work, we will consider the star network scenario where the central party is trusted while all the edge parties (with a number of $n$) are untrusted. Network steering is defined with an $n$ local hidden state model which can be viewed as a special kind of $n$ local hidden variable model. Two  different types of sufficient criteria,  nonlinear steering inequality and  linear steering inequality will be constructed to verify the quantum steering in a star network. Based on the linear steering inequality, how to detect the network steering with a fixed measurement will be discussed.
\end{abstract}

\pacs{03.65.Ud, 03.65.Ta}
%03.65.Ud Entanglement and quantum nonlocality
%03.65.Ta Foundations of quantum mechanics; measurement theory
\maketitle

\maketitle

\section{Introduction}
In 1930s, the concept of steering was introduced by Schr\"odinger~\cite{Sch} as a generalization of the Einstein-Podolsky-Rosen (EPR) paradox~\cite{Ein}. For a bipartite state, steering infers that an observer on one side can affect the state of the other spatially separated system by local measurements. In 2007, a standard formalism of quantum steering was developed by  Wiseman, Jones and Doherty~\cite{Wiseman1}.  In quantum information processing, EPR steering can
be defined as the task for a referee to determine whether one party
shares entanglement with a second untrusted party~\cite{Wiseman1,JWD,sau}.  Quantum steering  is a type of quantum nonlocality that is logically distinct from inseparability~\cite{Guhne,Horos} and Bell nonlocality~\cite{Brunner}.

In the last decade, the investigation of nonlocality has moved beyond Bell's theorem to consider more sophisticated experiments that involve several independent sources which distribute shares of physical systems among many parties in a network~\cite{Jones7,Jones8, Jones9}. The discussions, which are about the main concepts, methods, results and future challenges in the emerging topic of Bell in networks, can be found in the review article~\cite{TAV}. The independence of various sources leads to nonconvexity in the space of relevant correlations~\cite{Jones10,Jones11,Jones12,Jones13,Jones14,Jones15, Jones16,Jones17, Jones18}.

The simplest network scenario is provided by entanglement swapping~\cite{An16}. To contrast classical and quantum correlation in this scenario, the so-called bilocality assumption where the classical models consist of two independent local hidden variables (LHV), has been considered~\cite{Jones7, Jones8}. The generalization of the bilocality scenario to network, is the so-called $n$-locality scenario, where the number of independent sources of states is increased to arbitrary $n$~\cite{Jones11,Jones12,Jones13,An24,An26,An}. Some new interesting effects, such as the possibility to certify quantum nonlocality ``without inputs", are  offered by the network structure~\cite{Jones8,Jones9, Jones20,Jones21}.

The quantum network scenarios, where some of the parties are trusted while the others are untrusted,  are naturally connected to the notion of quantum steering. Though the notion of multipartite steering has been previously considered~\cite{he-reid,csan}, the steering in the scenario of network with independent sources is seldom discussed. In the recent work~\cite{Jonesi},  focusing on the linear network with trusted end points and intermediated untrusted parties who perform a fixed measurement, the authors introduced the network steering and network local hidden state models. Motivated by the work in Ref.~\cite{Jonesi}, the steering in a star network will be considered here.

An important example for a multiparty network is a star-shaped configuration. Such a star network is composed of a central party that is separately connected, via a number of $n$ independent bipartite sources, to $n$ edge parties. Correlations in the network arise through the central party jointly measuring the $n$ independent shares received from the sources and edge parties locally measuring the single shares received from the corresponding sources. In the present work, we consider the star network scenario where the central party is trusted while all the edge parties are untrusted.

In this work, the quantum steering in a star network is defined by introducing of an $n$-local hidden state (LHS) model. It will be shown that this $n$-LHS model can be viewed as a special kind of $n$-LHV model developed in~\cite{Jones11,Jones12,Jones13,An24,An26,An}. Besides the $n$-LHS model, we will focus on how to verify the quantum steering in the star network. Two different types of sufficient criteria,  linear steering inequality and  nonlinear steering inequality, will be designed.  Unlike detecting steering with single source, it will be shown that the network steering can be demonstrated with a fixed measurement  performed by the trusted central party.

The  content of this work is organized as follows. In Sec.~\ref{Sec2}, we give a brief review on the definition of steering in bipartite system. In Sec.~\ref{Sec3},   for  a star network scenario where the central party is trusted while all the edge parties are untrusted, network steering is defined with an $n$-LHS model. To verify the network steering,  the nonlinear steering inequality, linear steering inequality,  are designed in Sec.~\ref{Sec4}, and Sec.~\ref{Sec5}, respectively. Finally, we end our work with a short conclusion.

\section{Preliminary}
\label{Sec2}

\begin{figure}
\centering
\includegraphics[scale=0.2]{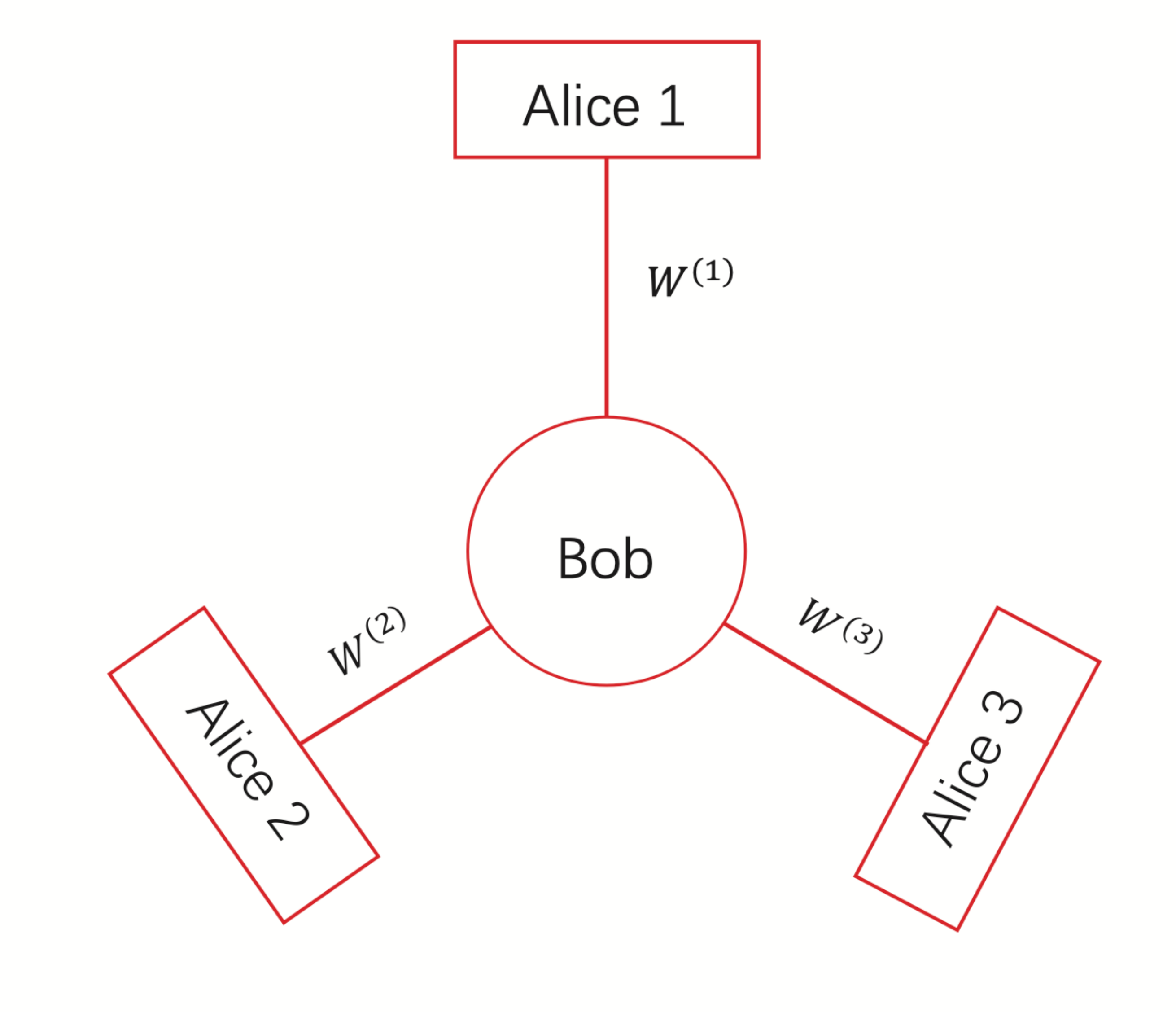}
\caption{A star network is composed of a central party and $n$ edge parties, and via a number of $n$ independent bipartite sources, the central party is separately connected to the $n$ edge parties. The figure above is depicted for a star-shaped network with $n=3$, where each edge observer shares a bipartite state $W^{(\mu)}$ $(\mu=1, 2, 3)$ with the central observer.}
\label{fig1}
\end{figure}

For convenience, in a star network shown in Fig.~\ref{fig1}, we call the observer in the central party as Bob while the observers in the edge parties as Alices.
Before defining quantum steering in the star network,  some necessary conventions are required. First, for the bipartite state $W^{(\mu)}$, the state shared between the the $\mu$th Alice and Bob,
 the $\mu$th Alice can perform $N$  measurements on her side, labelled by $x_{\mu}=1,2,...,N$, each having $m$ outcomes $a_{\mu}=0,1,...,m-1$, and the measurements are represented by $\hat{\Pi}^{(\mu)}_{a_{\mu}\vert x_{\mu}}$, $\sum_{a_\mu=0}^{m-1}\hat{\Pi}^{(\mu)}_{a_{\mu}\vert x_{\mu}}=I_d $, with $I_d$ the identity operator for the local $d$-dimensional Hilbert space. For a bipartite state $W^{(\mu)}$,  the unnormalized postmeasurement states (UPS) prepared for Bob are given by
\begin{equation}
\label{df}
\tilde{\rho}^{(\mu)}_{a_{\mu}\vert x_{\mu}}=\tr_\mathrm{A}\left[\left(\hat{\Pi}^{(\mu)}_{a_{\mu}\vert x_{\mu}}\otimes I_d\right)W^{(\mu)}\right].
\end{equation}
The set of the unnormalized states, $\left\{\tilde{\rho}^{(\mu)}_{a_{\mu}\vert x_{\mu}}\right\}$, is usually called an \emph{assemblage}.

In 2007, Wiseman, Jones and Doherty formally defined quantum steering as the possibility of remotely generating ensembles that could not be produced by an LHS model~\cite{Wiseman1}. An LHS model refers to the case where a source sends a classical message $\xi_{\mu}$ to the $\mu$th Alice, and a corresponding quantum state $\rho^{(\mu)}_{\xi_{\mu}}$ to Bob. Given that the $\mu$th Alice decides to perform the measurement $x_{\mu}$, the variable $\xi_{\mu}$ instructs the output $a_{\mu}$ of Alice's apparatus with the probability $\mathfrak{p}^{(\mu)}\left(a_{\mu}\vert x_{\mu},\xi_{\mu}\right)$. The variable $\xi_{\mu}$ can also be interpreted as an LHV and chosen according to a probability distribution $\Omega^{(\mu)}(\xi_{\mu})$. Bob does not have access to the classical variable $\xi_{\mu}$, and his final assemblage is composed by
\begin{equation}
\label{tilderho}
\tilde{\rho}^{(\mu)}_{a_{\mu}\vert x_{\mu}}=\int d\xi_{\mu}\Omega^{(\mu)}(\xi_{\mu})\mathfrak{p}^{(\mu)}(a_{\mu}\vert x_{\mu},\xi_{\mu})\rho^{(\mu)}_{\xi_{\mu}},
\end{equation}
with the constraints
\begin{equation}
\label{constraint}
\sum_{a_{\mu}} \mathfrak{p}^{(\mu)}(a_{\mu}\vert x_{\mu},\xi_{\mu})=1,\ \ \int d\xi_{\mu}\Omega^{(\mu)}(\xi_{\mu})=1.
\end{equation}

In this paper, the definition of steering from the $\mu$th Alice to Bob is directly cited from the review article~\cite{Can1}: An assemblage is said to demonstrate steering if it does not admit a decomposition of the form in Eq.~\eqref{tilderho}. Furthermore, a quantum state $W^{(\mu)}$ is said to be steerable from $\mu$th Alice to Bob if the experiments in $\mu$th Alice's part produce an assemblage that demonstrates steering. On the contrary, an assemblage is said to be LHS if it can be written as in Eq.~\eqref{tilderho}, and a quantum state is said to be unsteerable if an LHS assemblage is generated for all local measurements.

Joint measurability, which  is a natural extension of commutativity for general measurement, was studied extensively for a few decades before steering was formulated in its modern form~\cite{rmd}. Its relation with quantum steering has been discussed in recent works~\cite{quint, ula, UULA,Kiukas,Ku}. A set of measurements $\left\{\hat{M}^{(\mu)}_{ a_{\mu}\vert x_{\mu}}\right\}$, where $\sum_{a_{\mu}}\hat{M}^{(\mu)}_{ a_{\mu}\vert x_{\mu}}=I$ holds for each setting $x_{\mu}$,  is jointly measurable if there exists a set of
positive-operator-valued measures
 (POVMs) $\left\{\hat{M}^{(\mu)}_{\lambda}\right\}$, such that $\hat{M}^{(\mu)}_{a_\mu\vert x_\mu}=\sum_{\lambda}\pi(\lambda) p\left(a_{\mu}\vert x_{\mu},\lambda\right)\hat{M}_{\lambda}$ for all $a_{\mu}$ and $x_{\mu}$, with $\pi(\lambda)$ and  $p\left(a_{\mu}\vert x_{\mu},\lambda\right)$ the probability distributions. Otherwise, $\left\{\hat{M}^{(\mu)}_{ a_{\mu}\vert x_{\mu}}\right\}$ is said to be incompatible.

For the bipartite state $W^{(\mu)}$ shared by Bob and the $\mu$th Alice, one can denote $\rho_\mathrm{B}$ to be the reduced density matrix on Bob's side and the purification of $\rho_{\mathrm{B}}^*$ is denoted by $\vert\Psi^{(\mu)}\rangle$, where $\rho_{\mathrm{B}}^*$ is the complex conjugate of $\rho_{\mathrm{B}}$. The state $W^{(\mu)}$ can be expressed as
\begin{equation}
W^{(\mu)}=\varepsilon\otimes \mathcal{I}\left(\vert\Psi^{(\mu)}\rangle\langle \Psi^{(\mu)}\vert\right),
\nonumber
\end{equation}
by introducing a quantum channel $\varepsilon$, $\varepsilon(\rho)=\sum_m \hat{E}_m\rho \hat{E}^{\dagger}_m$ with the Kraus operators $\left\{\hat{E}_m\right\}$, and $\mathcal{I}$ is an identity map.
With the definition that $\hat{M}^{(\mu)}_{a_{\mu}\vert x_{\mu}}=\sum_m \hat{E}_{m}^{\dagger}\hat{\Pi}^{(\mu)}_{a_{\mu}\vert x_{\mu}}\hat{E}_m$, it has been shown that the conditional states in Eq.~\eqref{df}
can be reexpressed as
\begin{equation}
\label{jm=lhs}
\tilde{\rho}^{(\mu)}_{a_{\mu}\vert x_{\mu}}=\rho^*_\mathrm{B}\hat{M}^{(\mu)}_{a_{\mu}\vert x_{\mu}}\left(\rho^*_\mathrm{B}\right)^{\dagger}.
\end{equation}
If $\left\{\tilde{\rho}^{(\mu)}_{a_{\mu}\vert x_{\mu}}\right\}$ admits an LHS model, by an explicit definition of the inverse matrix of $\rho_\mathrm{B}$, one may find that the measurement $\left\{\hat{M}^{(\mu)}_{a_{\mu}\vert x_{\mu}}\right\}$ is jointly measurable~\cite{rmd,ula}.

For the state $W^{(\mu)}$,  the measurement performed by the $\mu$th Alice is denoted by $\hat{\Pi}^{(\mu)}_{a_{\mu}\vert x_{\mu}}$, and if each $a_{\mu}$ takes the value 0 or 1, one can introduce an operator
\begin{equation}
\label{dfA}
\hat{A}^{(\mu)}_{x_{\mu}}=\hat{\Pi}^{(\mu)}_{0\vert x_{\mu}}-\hat{\Pi}^{(\mu)}_{1\vert x_{\mu}}.
\end{equation}
Similarly, the measurement performed by Bob is denoted by $\hat{\Pi}^{(\mu)}_{b_{\mu}\vert y_{\mu}}$, and if $b_{\mu}$ takes the value 0 or 1, another operator $\hat{B}_{y_{\mu}}$ can also be introduced
\begin{equation}
\label{dfB}
\hat{B}^{(\mu)}_{y_{\mu}}=\hat{\Pi}^{(\mu)}_{0\vert y_{\mu}}-\hat{\Pi}^{(\mu)}_{1\vert y_{\mu}}.\nonumber
\end{equation}

For the case $n=2$, let $ \hat{O}=\left(\hat{A}^{(1)}_{x_1}\otimes\hat{B}^{(1)}_{y_1}\right)\otimes\left(\hat{A}^{(2)}_{x_2}\otimes\hat{B}^{(2)}_{y_2}\right)$ to be an observable for the state $W=\otimes_{\mu=1}^2W^{(\mu)}$. Formally, one can introduce a unitary transformation  $U$, which is defined by $\hat{O}'\equiv U\hat{O}U^{\dagger}=\left(\hat{A}^{(1)}_{x_1}\otimes\hat{A}^{(2)}_{x_2}\right)\otimes\left(\hat{B}^{(1)}_{y_1}\otimes\hat{B}^{(2)}_{y_2}\right)$ and $W'=U^{\dagger}W U$, to express the expectation $\tr\left(\hat{O}W\right)$ as $\left\langle \hat{O}'\right\rangle\equiv\tr\left(\hat{O}'W'\right)$. Certainly, the above definition can be easily generalized for arbitrary $n$.

\section{$n\rightarrow 1$ steering in a star network}
\label{Sec3}

The concept of $n$-locality has been considered in the star-network configuration. Let $\lambda_{\mu}$ be LHV for the $\mu$th resource $W^{(\mu)}$, $\int d\lambda_{\mu}\omega^{(\mu)}(\lambda_{\mu})=1$, and $\bar{\lambda}=\lambda_1\lambda_2...\lambda_n$. If
\begin{equation}
\label{n-lhv}
p(\bar{a},b\vert \bar{x},y)=\int d\bar{\lambda}\left[\prod_{\mu=1}^n \omega^{(\mu)}(\lambda_{\mu}){p}^{(\mu)}(a_{\mu}\vert x_{\mu},\lambda_{\mu})\right]
{p}(b\vert y, \bar{\lambda}),
\end{equation}
with $d\bar{\lambda}=d\lambda_1d\lambda_2...d\lambda_\mu...d\lambda_n$, where ${p}^{(\mu)}(a_{\mu}\vert x_{\mu},\lambda_{\mu})$ and ${p}(b\vert y, \bar{\lambda})$ are the predetermined values of the operators $\hat{\Pi}^{(\mu)}_{a_{\mu}\vert x_{\mu}}$ and $\hat{\Pi}^b_y$ within the $n$-LHV model, respectively, the probabilities $p(\bar{a},b\vert \bar{x},y)$ admit the $n$-LHV model~\cite{Jones11,Jones12,Jones13,An24,An26,An}. Otherwise, if the set of correlations $\left\{p(\bar{a},b\vert\bar{x},y)\right\}$ does not admit the $n$-LHV model, it is said to be $n$-nonlocal. 

In above section, for a bipartite state $W^{(\mu)}$, the UPS $\tilde{\rho}^{(\mu)}_{a_{\mu}\vert x_{\mu}}$ and LHS $\rho^{(\mu)}_{\xi_{\mu}}$ have been given in Eq.~\eqref{df} and Eq.~\eqref{tilderho}, respectively. In the present work, the state in the star network can be expressed as
\begin{equation}
\label{definingW}
W=\bigotimes_{\mu=1}^n W^{(\mu)}.
\end{equation}
Furthermore, it is supposed that $\mu$th Alice can perform $N^{\mu}$ local measurements on her side, labelled by $x_{\mu}=1,2,...,N^{\mu}$, each having $m$ outcomes $a_{\mu}=0,1,...,m-1$, and the measurements are represented by $\hat{\Pi}^{(\mu)}_{a_{\mu}\vert x_{\mu}}$, $\sum_{a_\mu=0}^{m-1}\hat{\Pi}^{(\mu)}_{a_{\mu}\vert x_{\mu}}=I $.	With the state $W$ and the set of local measurements  $\left\{\bigotimes_{\mu=1}^n\hat{\Pi}^{(\mu)}_{a_{\mu}\vert x_{\mu}}\right\}$, one may have an assemblage of the conditional states $\left\{\bigotimes^{n}_{\mu=1}\tilde{\rho}^{(\mu)}_{a_{\mu}\vert x_{\mu}}\right\}$ on Bob's side. Now, one can introduce the following definition: If all the states in the assemblage can be expanded as
\begin{equation}
\label{tildenrho}
\bigotimes^{n}_{\mu=1}\tilde{\rho}^{(\mu)}_{a_{\mu}\vert x_{\mu}}=\bigotimes^{n}_{\mu=1}\int d\xi_{\mu}\Omega^{(\mu)}(\xi_{\mu})\mathfrak{p}^{(\mu)}(a_{\mu}\vert x_{\mu},\xi_{\mu})\rho^{(\mu)}_{\xi_{\mu}},
\end{equation}
the assemblage admits  an $n$-LHS model.  On the other hands, if the assembalge does not admit such an  $n$-LHS model, one may conclude that the state $W$ is steerable from Alices to Bob. Unlike the standard definition of steering for a bipartite state, in the star-network,  there are $n$ edge observers and one trusted central party. Therefore, we call the steering defined above as $n\rightarrow 1$ steerability. Finally, if   all the possible local measurements $\left\{\bigotimes_{\mu=1}^n\hat{\Pi}^{(\mu)}_{a_{\mu}\vert x_{\mu}}\right\}$ performed by the edge observers have been considered, and the assemblage of the conditional states $\left\{\bigotimes^{n}_{\mu=1}\tilde{\rho}^{(\mu)}_{a_{\mu}\vert x_{\mu}}\right\}$ always admits an $n$-LHS model, we say that the state $W$ is $n\rightarrow 1$ unsteerable. 

If each  resource state is unsteerable to Bob, for all the possible local measurements $\left\{\bigotimes_{\mu=1}^n\hat{\Pi}^{(\mu)}_{a_{\mu}\vert x_{\mu}}\right\}$ performed by the edge observers, the assemblage of the conditional states $\left\{\bigotimes^{n}_{\mu=1}\tilde{\rho}^{(\mu)}_{a_{\mu}\vert x_{\mu}}\right\}$ should admit an $n$-LHS model, and therefore, for the product state in Eq.~\eqref{definingW}, the detection of $n\rightarrow 1$ steering indicates that  there exists at least one source state $W^{(\mu)}$ which is one-way steerable from the $\mu$th Alice to Bob.

One of the main reasons for us to introduce the $n\rightarrow 1$ steering  is that it represents a weak form of the $n$-nonlocality. In other words, if the state $W$ is $n$-nonlocal, it must be $n\rightarrow1$ steerable. It can be easily verified that  the $n$-LHS model can be viewed as a special kind of $n$-LHV model. Formally, the POVMs performed by Bob can be denoted by $\hat{\Pi}^b_y$, $\sum_b\hat{\Pi}^b_y=I$, and the denotations $\bar{a}=a_1a_2...a_n$, $\bar{x}=x_1x_1...x_n$, and $\bar{\xi}=\xi_1\xi_2...\xi_n$ can be introduced.
   In the star-shapped network,  for the trusted party, the quantum mechanics is allowed, say, $\mathfrak{p}(b\vert y, \bar{\xi})=\tr\left[\hat{\Pi}^b_y\left(\bigotimes^{n}_{\mu=1}{\rho}^{(\mu)}_{\xi_{\mu}}\right)\right]$. 
   Within the $n$-LHS model,  the expectation $p(\bar{a},b\vert \bar{x},y)$ of the operator $\left(\bigotimes_{\mu=1}^n\hat{\Pi}^{(\mu)}_{a_{\mu}\vert x_{\mu}}\right)\otimes\hat{\Pi}^b_y$ can be expressed as
 \begin{eqnarray}
 &&p(\bar{a},b\vert \bar{x},y)=\int d\bar{\xi}\left[\prod_{\mu=1}^n \Omega^{(\mu)}(\xi_{\mu})\mathfrak{p}^{(\mu)}(a_{\mu}\vert x_{\mu},\xi_{\mu})\right]\mathfrak{p}(b\vert y, \bar{\xi}),\nonumber\\
 \label{statistics}
 &&\mathfrak{p}(b\vert y, \bar{\xi})=\tr\left[\hat{\Pi}^b_y\left(\bigotimes^{n}_{\mu=1}\tilde{\rho}^{(\mu)}_{\xi_{\mu}}\right)\right],
 \end{eqnarray}
 where $d\bar{\xi}=d\xi_1d\xi_2...d\xi_{\mu}...d\xi_{n}$.  Obviously, it belongs to the $n$-LHV model in Eq.~\eqref{n-lhv}.

In the following, we will focus on the case where each independent source state is a two-qubit state and develop several types of sufficient criteria to verify the $n\rightarrow1$ steering.

\section{nonlinear steering inequality}
\label{Sec4}
For $n=1$, there is a well-known nonlinear steering inequality~\cite{Can3}
\begin{eqnarray}
\label{Bell-like}
&&\sqrt{\left\langle\left(\hat{A}_1+\hat{A}_2\right)\bigotimes \hat{B}_1\right\rangle^2
+\left\langle\left(\hat{A}_1+\hat{A}_2\right)\bigotimes \hat{B}_2\right\rangle^2}\nonumber\\
&&+\sqrt{\left\langle\left(\hat{A}_1-\hat{A}_2\right)\bigotimes \hat{B}_1\right\rangle^2
+\left\langle\left(\hat{A}_1-\hat{A}_2\right)\bigotimes \hat{B}_2\right\rangle^2}\leq 2,\nonumber\\
\end{eqnarray}
where mutually unbiased measurements are performed on Bob's site~\cite{Can3}. This inequality has a peculiar property: If the state  violates the original
Clauser-Horne-Shimony-Holt (CHSH) inequality~\cite{chsh}, it must violate the inequality above. Next, an inequality can be constructed to verify the $n\rightarrow1$ steering, and this inequality can be viewed as a generalization of the one in Eq.~\eqref{Bell-like} above.

Let $\sigma^{(\mu)}_i$, with $i=1, 2, 3$ be the Pauli matrices for the $\mu$th source state $W^{(\mu)}$. $\bm{r}^{(\mu)}_{x_{\mu}}$ is a three-dimensional vector in $\mathbb{R}^3$, $\bm{r}^{(\mu)}_{x_{\mu}}=\left(r^{(\mu)}_{x_{\mu}1},r^{(\mu)}_{x_{\mu}2},r^{(\mu)}_{x_{\mu}3}\right)^{\mathrm{T}}$, and the Euclidean norm of $\bm{r}^{(\mu)}_{x_{\mu}}$ is denoted by $\left\vert\left\vert\bm{r}^{(\mu)}_{x_{\mu}}\right\vert\right\vert\equiv\sqrt{\sum_{j=1}^3\left(r^{(\mu)}_{x_{\mu}j}\right)^2}$. Now, consider the case that the experiment setting for each Alice is two, $x_{\mu}\in\{1,2\}$, $\forall\mu\in\{1,2,...,n\}$, and the POVMs $\hat{M}^{(\mu)}_{a_{\mu}\vert x_{\mu}}$ $(a_{\mu}\in\{0,1\}$) may take the general form
\begin{equation}
\label{POVMS}
\hat{M}^{(\mu)}_{0\vert x_{\mu}}=\frac{1}{2}\left[(1+k_{x_{\mu}})I+ \bm{r}^{(\mu)}_{x_{\mu}}\cdot\bm{\sigma}\right],\ \
\hat{M}^{(\mu)}_{1\vert x_{\mu}}=I-\hat{M}^{(\mu)}_{0\vert x_{\mu}},
\end{equation}
with $-1\leq k_{x_{\mu}}\leq 1$ and $\left\vert\left\vert\bm{r}^{(\mu)}_{x_{\mu}}\right\vert\right\vert\leq1+k_{x_{\mu}}$. As a well-known result~\cite{rmd,Pusey}, the necessary condition for the pair of POVMs $\left\{\hat{M}^{(\mu)}_{0\vert 1},\hat{M}^{(\mu)}_{1\vert 1}\right\}$ and $\left\{\hat{M}^{(\mu)}_{0\vert 2},\hat{M}^{(\mu)}_{1\vert 2}\right\}$ to be jointly measurable is
\begin{equation}
\label{pair}
\left\vert\left\vert\bm{r}^{(\mu)}_{1}+\bm{r}^{(\mu)}_{2}\right\vert\right\vert+
\left\vert\left\vert\bm{r}^{(\mu)}_{1}-\bm{r}^{(\mu)}_{2}\right\vert\right\vert\leq 2.
\end{equation}

For the POVMs in Eq.~\eqref{POVMS}, one can redefine the operator $\hat{A}^{(\mu)}_{x_{\mu}}=\hat{M}^{(\mu)}_{0\vert x_{\mu}}-\hat{M}^{(\mu)}_{1\vert x_{\mu}}$, which is not traceless in general,
\begin{equation}
\label{operatorA}
\hat{A}^{(\mu)}_{x_{\mu}}=k_{x_{\mu}}I+ \bm{r}^{(\mu)}_{x_{\mu}}\cdot\bm{\sigma}^{(\mu)}.
\end{equation}
With the fact that any two-qubit state $\rho_{\mathrm{AB}}$  can be decomposed as
\begin{eqnarray}
\rho_{\mathrm{AB}}&=&
\frac{1}{4}\bigg[I\otimes I+\bm{r}_{\mathrm{A}}\cdot\bm{\sigma}\otimes I \nonumber\\
&+& I\otimes \bm{r}_{\mathrm{B}}\cdot\bm{\sigma}+\sum_{i,j=1}^3 t_{ij}\sigma_i\otimes\sigma_j\bigg],
\end{eqnarray}
the two reduced density matrices can be expressed as $\rho_{\mathrm{A}}=(I+\bm{r}_{\mathrm{A}}\cdot\bm{\sigma})/2$ and $\rho_{\mathrm{B}}=(I+\bm{r}_{\mathrm{B}}\cdot\bm{\sigma})/2$. The coefficients $t_{ij}=\tr\left[\rho_{\mathrm{AB}}(\sigma_i\otimes\sigma_j)\right]$ form a real matrix (which is referred as the T-matrix) denote by $T_{\rho}$~\cite{HHH}. With bracket $(\bullet,\bullet)$ standing for Euclidean scalar product of two vectors in $\mathbb{R}^3$, the expectation $\langle \bm{r}\cdot\bm{\sigma}\otimes \bm{n}\cdot\bm{\sigma}\rangle$ can be expressed as $\langle \bm{r}\cdot\bm{\sigma}\otimes \bm{n}\cdot\bm{\sigma}\rangle=
(\bm{n},T^{\mathrm{T}}_{\rho}\bm{r})$, where $T^{\mathrm{T}}_{\rho}$ is the transpose of $T_{\rho}$.

If $\tr\hat{A}^{(\mu)}_{x_{\mu}}\neq 0$, setting $\hat{B}^{(\mu)}_{y_{\mu}}=\bm{n}^{(\mu)}_{y_{\mu}}\cdot \bm{\sigma}^{(\mu)}$ with $\left\vert\left\vert \bm{n}^{(\mu)}_{y_{\mu}}\right\vert\right\vert=1 $, and  introducing the following constraint
\begin{equation}
\label{constraint}
\tr\left(\hat{B}^{(\mu)}_{y_{\mu}}\rho_{\mathrm{B}}\right)=0,
\end{equation}
one can easily verify that
\begin{equation}
\label{expectation}
\left\langle\hat{A}^{(\mu)}_{x_{\mu}}\bigotimes\hat{B}^{(\mu)}_{y_{\mu}}\right\rangle=\left\langle\left(\bm{r}^{(\mu)}_{x_{\mu}}\cdot\bm{\sigma}^{(\mu)}\right)\bigotimes\hat{B}^{(\mu)}_{y_{\mu}}\right\rangle.
\end{equation}
With $p^{(\mu)}(b_{\mu}\vert y_{\mu})=\tr\left[\hat{\Pi}^{(\mu)}_{b_{\mu}\vert y_{\mu}}\rho_{\mathrm{B}}\right]$, Eq.~\eqref{constraint} is equivalent to
\begin{equation}
p^{(\mu)}({0}\vert y_{\mu})=p^{(\mu)}({1}\vert y_{\mu})=\frac{1}{2}.
\end{equation}

Now, a pair of vectors $\bm{s}^{(\mu)}$ and $\bm{t}^{(\mu)}$ can be introduced for a fixed $\mu$, which satisfy the conditions $\left\vert\left\vert\bm{s}^{(\mu)}\right\vert\right\vert\leq 1$ and  $\left\vert\left\vert\bm{t}^{(\mu)}\right\vert\right\vert\leq 1$. The necessary condition in Eq.~\eqref{pair} can be equivalently expressed as
\begin{eqnarray}
\label{equivalent}
\frac{1}{2}\left(\bm{r}^{(\mu)}_{1}+\bm{r}^{(\mu)}_{2}\right)&=&\cos^{2}\omega_{\mu}\bm{s}^{(\mu)},\\
\frac{1}{2}\left(\bm{r}^{(\mu)}_{1}-\bm{r}^{(\mu)}_{2}\right)&=&\sin^{2}\omega_{\mu}\bm{t}^{(\mu)}.
\end{eqnarray}
With $T_{\mu}$ the T-matrix of the state $W^{(\mu)}$, one can have
\begin{eqnarray}
\label{exp21}
\left\langle\frac{1}{2}\left(\hat{A}^{(\mu)}_1+\hat{A}^{(\mu)}_2\right)\otimes \hat{B}^{(\mu)}_{y_{\mu}}\right\rangle&=&\cos^2\omega_{\mu}\left(\bm{n}^{(\mu)}_{y_{\mu}},T_{\mu}^{\mathrm{T}}\bm{s}^{(\mu)}\right),\\
\label{exp22}
\left\langle\frac{1}{2}\left(\hat{A}^{(\mu)}_1-\hat{A}^{(\mu)}_2\right)\otimes \hat{B}^{(\mu)}_{y_{\mu}}\right\rangle&=&\sin^2\omega_{\mu}\left(\bm{n}^{(\mu)}_{y_{\mu}},T_{\mu}^{\mathrm{T}}\bm{t}^{(\mu)}\right).
\end{eqnarray}

Formally, using the parameters $\alpha_{\mu}\in\{0,1\}$, one can define the operators
\begin{equation}
\hat{A}_{\alpha_1\alpha_2...\alpha_n}=\frac{1}{2}\bigotimes_{\mu=1}^n\left(\hat{A}^{(\mu)}_1+(-1)^{\alpha_{\mu}}\hat{A}^{(\mu)}_2\right).
\end{equation}
Let the experimental setting $y_{\mu}$ be $y_{\mu}=1+\alpha_{\mu}$, the operators for Bob are defined as
\begin{equation}
\label{BobB}
\hat{B}_{\alpha_1\alpha_2...\alpha_n}=\bigotimes_{\mu=1}^n \left(\bm{n}^{(\mu)}_{\alpha_{\mu}+1}\cdot\bm{\sigma}^{(\mu)}\right).
\end{equation}
Besides the constraint in Eq.~\eqref{constraint}, it is required that $\left(\bm{n}^{(\mu)}_1,\bm{n}^{(\mu)}_2\right)=0$, for all $\mu\in\{1,2,...,n\}$. Furthermore, introducing the denotations $\bar{0}=00...00$ (corresponding to each $\alpha_{\mu}=0$) and $\bar{1}=11...1$ (corresponding to each $\alpha_{\mu}=1$), a non-linear inequality can be obtained (for the details, please see the Appendix),
\begin{eqnarray}
\label{non-linear}
&&\sqrt{\left(\left\langle\hat{A}_{\bar{0}}\bigotimes\hat{B}_{\bar{0}}\right\rangle^2\right)^{\frac{1}{n}}
+\left(\left\langle\hat{A}_{\bar{0}}\bigotimes\hat{B}_{\bar{1}}\right\rangle^2\right)^{\frac{1}{n}}}\nonumber\\
&&+\sqrt{\left(\left\langle\hat{A}_{\bar{1}}\bigotimes\hat{B}_{\bar{0}}\right\rangle^2\right)^{\frac{1}{n}}
+\left(\left\langle\hat{A}_{\bar{1}}\bigotimes\hat{B}_{\bar{1}}\right\rangle^2\right)^{\frac{1}{n}}}\leq 1,
\end{eqnarray}
and this is the necessary condition that the measurements performed by each Alice are jointly measurable. According to the general relation between the compatible measurement and the LHS model in Eq.~\eqref{jm=lhs}, the above inequality is also a necessary condition that the set of probabilities $\{p(\bar{a}\vert \bar{x},b\vert y)\}$ admits the $n$-LHS model defined in Eq.~\eqref{statistics}. Therefore, if it is violated, one can conclude that the state $W$ is $n\rightarrow1$ steerable. As expected, if $n=1$, the inequality in Eq.~\eqref{Bell-like} is recovered. From the inequality in Eq.~\eqref{non-linear}, one can obtain a more simplified inequality,
\begin{equation}
\left\vert\left\langle\hat{A}_{\bar{0}}\bigotimes\hat{B}_{\bar{0}}\right\rangle\right\vert^{\frac{1}{n}}
+\left\vert\left\langle\hat{A}_{\bar{1}}\bigotimes\hat{B}_{\bar{1}}\right\rangle\right\vert^{\frac{1}{n}}\leq 1.
\end{equation}
which is similar to the well-known criterion to verify the $n$-nonlocality~\cite{Jones11}. The inequality above is equivalent to the original one iff $\left\langle\hat{A}_{\bar{0}}\otimes\hat{B}_{\bar{1}}\right\rangle =\left\langle\hat{A}_{\bar{1}}\otimes\hat{B}_{\bar{0}}\right\rangle=0.$

Assuming that each source state $W^{(\mu)}$ is a maximally entangled state, and choosing the measurements performed by the $\mu$th Alice as
\begin{equation}
\label{Asetting}
\hat{A}^{(\mu)}_1+\hat{A}^{(\mu)}_2={\sqrt{2}}\sigma^{(\mu)}_1,\ \ 
\hat{A}^{(\mu)}_1-\hat{A}^{(\mu)}_2={\sqrt{2}}\sigma^{(\mu)}_3
\end{equation}
and the measurements performed by Bob as
\begin{equation}
\hat{B}_{\bar{0}}=\bigotimes_{\mu=1}^n \sigma^{(\mu)}_1,\ \
\hat{B}_{\bar{1}}=\bigotimes_{\mu=1}^n \sigma^{(\mu)}_3,
\end{equation}
there should be $\left(\left\langle\hat{A}_{\bar{0}}\otimes\hat{B}_{\bar{0}}\right\rangle^2\right)^{1/n}=\left(\left\langle\hat{A}_{\bar{1}}\otimes\hat{B}_{\bar{1}}\right\rangle^2\right)^{1/n}=1/2$ and $\left(\left\langle\hat{A}_{\bar{0}}\otimes\hat{B}_{\bar{1}}\right\rangle^2\right)^{1/n}=\left(\left\langle\hat{A}_{\bar{1}}\otimes\hat{B}_{\bar{0}}\right\rangle^2\right)^{1/n}=0$. Under such choices, the inequality in Eq.~\eqref{non-linear} is violated by a factor $\sqrt{2}$, which is independent of $n$.

\section{Linear steering inequality}
\label{Sec5}
In the star network, Bob can perform joint measurement on all the particles in his hand. It has been already  shown that the $n$-nonlocality  can be detected even though a fixed projective measurement is performed by Bob. Certainy, a fixed measurment can be applied to detecting  the steering. In follwing, we shall at first derive a linear inequality and then shown how to apply it with a fixed measurement.

Assume that Bob has two spin-1/2 particles, and with the four maximally entangled states,
\begin{equation}
\label{BSM}
\vert\Psi^{\pm}\rangle=\frac{1}{\sqrt{2}}(\vert00\rangle\pm\vert11\rangle),\ \
\vert\Phi^{\pm}\rangle=\frac{1}{\sqrt{2}}(\vert01\rangle\pm\vert10\rangle),
\end{equation}
the standard Bell measurement (SBM) consists of four rank-one operators $\left\{\hat{\Psi}^{\pm},\hat{\Phi}^{\pm}\right\}$, where the single capital letter $\hat{\Psi}$ stands for $\vert\Psi\rangle\langle\Psi\vert$. In the following, an explicit example is given to show that the $2\rightarrow1$ can be verified with the SBM.

The way of constructing linear steering inequalities (LSIs) originates from the works in Refs.~\cite{can22,sau,Joness}. For a single bipartite system, to discuss the one-way steering from Alice (the untrusted party) to Bob (the trusted party), one may construct a criterion which only depends on the measurements performed by Bob. Besides the property that the LSIs can work even when the state is unknown, they also have a deep relation with the compatible measurement: If a one-way LSI is violated, the state is steerable from Alice to Bob and the measurements performed by Alice are also verified to be incompatible~\cite{quint,ula,UULA,Kiukas,Wu,WU2,WU3}. Now, consider the star network with $n=2$, for the set of measurements performed by the $\mu$th Alice $\left\{\hat{\Pi}^{(\mu)}_{a_{\mu}\vert x_{\mu}}\right\}$, the number of the experimental settings is fixed to be three, $x_{\mu}\in\{1,2,3\}$, and for a given setting $x_{\mu}$, there are two measurement results, $a_{\mu}\in\{0,1\}$. One can introduce an operator
 \begin{equation}
 \label{threepauli}
 \hat{B}_j=\sigma^{(1)}_j\bigotimes \sigma^{(2)}_j,\ \ j=1,2,3,
 \end{equation}
where $\sigma^{(\mu)}_j$ ($\mu\in\{1,2\}$) are the Pauli matrices for the $\mu$th source state $W^{(\mu)}$. From the definition in Eq. \eqref{dfA}, another operator $\hat{H}$ can be introduced, which is defined as $\hat{H}=\sum_{j=1}^3\hat{A}^{(1)}_j\otimes \hat{A} ^{(2)}_j\otimes \hat{B}_j$. Our task is to find the maximum value of the expectation $\langle\hat{H}\rangle\equiv \tr\left(\tilde{W }\hat{H}\right)$ under the LHS model in Eq.~\eqref{tildenrho} with $n=2$,
\begin{eqnarray}
p(\bar{a},b\vert \bar{x},y)&=&\int d\xi_{1}\Omega^{1}(\xi_{1})\int d\xi_{2}\Omega^{2}(\xi_{2})\nonumber\\
\label{2lhs}
&&\times\mathfrak{p}^{(1)}(a_{1}\vert x_{1},\xi_{1})\mathfrak{p}^{(2)}(a_{2}\vert x_{2},\xi_{2})
\mathfrak{p}(b\vert y, \bar{\xi}),
\end{eqnarray}
where $\mathfrak{p}(b\vert y,\bar{\xi})=\mathrm{Tr}\left[\hat{\Pi}^b_y\left(\otimes^{2}_{\mu=1}\tilde{\rho}^{(\mu)}_{\xi_{\mu}}\right)\right]$. Here, the operators in Eq.~\eqref{threepauli} can always be expanded as $\hat{B}_j=\hat{\Pi}^0_j-\hat{\Pi}^1_j$, with $\hat{\Pi}^0_j+\hat{\Pi}^1_j=I$. To calculate the expectation $\langle\hat{H}\rangle$ within the 2-LHS model above, it is convenient to introduce a quantity $\mathfrak{d}^{(\mu)}(x_{\mu},\xi_{\mu})$,
\begin{equation}
\mathfrak{d}^{(\mu)}(x_{\mu},\xi_{\mu})=\mathfrak{p}^{(\mu)}(0\vert x_{\mu},\xi_{\mu})
-\mathfrak{p}^{(\mu)}(1\vert x_{\mu},\xi_{\mu}),\nonumber
\end{equation}
with $\mu\in\{1,2\}$ and $x_{\mu}\in\{1,2,3\}$. From the constraint $\mathfrak{p}^{(\mu)}(0\vert x_{\mu},\xi_{\mu})+\mathfrak{p}^{(\mu)}(1\vert x_{\mu},\xi_{\mu})=1$, there is
\begin{equation}
\label{constraintd}
-1\leq\mathfrak{d}^{(\mu)}(x_{\mu},\xi_{\mu})\leq 1.
\end{equation}
Now, one can introduce an operator $\hat{H}_{\xi_1\xi_2}$,
\begin{equation}
\label{dfH}
\hat{H}_{\xi_1\xi_2}=\sum_{j=1}^3 \mathfrak{d}^{(1)}(j,\xi_{1})\mathfrak{d}^{(2)}(j,\xi_{2})\hat{B}_j,
\end{equation}
and obtain
\begin{equation}
\langle\hat{H}\rangle=\int d\xi_{1}\Omega^{1}(\xi_{1})\int d\xi_{2}\Omega^{2}(\xi_{2})\tr\left[\left(\rho^{(1)}_{\xi_{1}}\otimes \rho^{(2)}_{\xi_{2}}\right)\hat{H}_{\xi_1\xi_2}\right],
\label{exp2lhs}
\end{equation}
where $\rho^{(\mu)}_{\xi_{\mu}}$ $(\mu=1,2)$ have been defined in Eq. \eqref{tilderho}. Finally, a quantity $\beta$ can be defined
\begin{equation}
\label{beta}
\beta=\max_{\{\vert\psi\rangle,\vert\phi\rangle\}}\langle \psi\vert\otimes\langle
\phi\vert \hat{H}_{\xi_1\xi_2}\vert\psi\rangle\otimes\vert\phi\rangle,
\end{equation}
with $\vert\psi\rangle\otimes\vert\phi\rangle$ an arbitrary pure product state, and it can be easily verified that $\tr\left[\left(\rho^{(1)}_{\xi_{1}}\otimes \rho^{(2)}_{\xi_{2}}\right)\hat{H}_{\xi_1\xi_2}\right]\leq \beta$. Based on the elementary relations that $\int d\xi_{1}\Omega^{1}(\xi_{1})=\int d\xi_{2}\Omega^{2}(\xi_{2})=1$, a LSI can be known as
\begin{equation}
\label{lsi22}
\langle \hat{H}\rangle\leq \beta.
\end{equation}
If the above inequality is violated, one can conclude that the state $W=\otimes_{\mu=1}^2 W^{(\mu)}$ is $2\rightarrow 1$ steerable.

Putting the operators defined in Eq.~\eqref{threepauli} into Eq.~\eqref{dfH}, the value of $\beta$ can be derived in a simple way. First, the pure states $\vert\psi\rangle$ and $\vert\phi\rangle$ can be described with their corresponding unit vectors in $\mathbb{R}^3$, say $\vert\psi\rangle\langle \psi\vert=(I+\bm{s}\cdot\bm{\sigma})/2$ and  $\vert\phi\rangle\langle\phi\vert=(I+\bm{t}\cdot\bm{\sigma})/2$. Then, with $s_i=\tr\left(\sigma_i\vert\psi\rangle\langle \psi\vert\right)$ and $t_i=\tr\left(\sigma_i\vert\phi\rangle\langle \phi\vert\right)$, where $i=1, 2, 3$, the two vectors can be expressed with their components as $\bm{s}=(s_1,s_2,s_3)^{\mathrm{T}}$ and $\bm{t}=(t_1,t_2,t_3)^{\mathrm{T}}$. With the denotations introduced above, two vectors can be defined
\begin{eqnarray}
\bm{s}'&=&\left(s_1 \mathfrak{d}^{(1)}(1,\xi_1), s_2 \mathfrak{d}^{(1)}(2,\xi_1), s_3 \mathfrak{d}^{(1)}(3,\xi_1)\right)^{\mathrm{T}},\nonumber\\
\bm{t}'&=&\left(t_1 \mathfrak{d}^{(2)}(1,\xi_2), t_2 \mathfrak{d}^{(2)}(2,\xi_2), t_3 \mathfrak{d}^{(2)}(3,\xi_2)\right)^{\mathrm{T}},\nonumber
\end{eqnarray}
which satisfy $\vert\vert \bm{s}'\vert\vert\leq \vert\vert\bm{s}\vert\vert=1$ and $\vert\vert \bm{t}'\vert\vert\leq\vert\vert\bm{t}\vert\vert=1$ according to Eq.~\eqref{constraintd}. $\beta$ can be expressed as $\beta=\max_{\bm{s},\bm{t}}( \bm{t}',\bm{s}')$, and certainly, $\beta=1$.

Now, we will consider a simple case $W^{(1)}=W^{(2)}=\rho$, where $\rho$ is an isotopic state for two-qubit system
\begin{equation}
\label{isotopic}
\rho=\eta\vert\Psi^{+}\rangle\langle\Psi^{+}\vert+\frac{1}{4}(1-\eta)I\otimes I.
\end{equation}
The measurements performed by each Alice are fixed as
\begin{equation}
\label{eachalice}
\hat{\Pi}^{(\mu)}_{a_{\mu}\vert x_{\mu}}=\frac{1}{2}\left[I+(-1)^{a_{\mu}}\sigma^{(\mu)}_{x_{\mu}}\right],
\end{equation}
with $\mu\in\{1,2\}$, $a_{\mu}\in\{0,1\}$ and $x_{\mu}\in\{1,2,3\}$. Using Eq.~\eqref{dfA} and Eq.~\eqref{threepauli}, it can be easily calculated that $\langle\hat{H}\rangle=3\eta^2$. Therefore, the maximum violation of the LSI $\langle \hat{H}\rangle\leq \beta\equiv1$ is attained when $\langle \hat{H}\rangle=3$ (since $\langle\hat{H}\rangle/\beta=3$ if $\eta=1$). If $\eta\leq\sqrt{3}/3$, it is known the set of conditional states, which are resulted from the measurements in Eq.~\eqref{eachalice}, should admit an LHS model. Now, the assemblage $\left\{\rho^{(1)}_{a_1\vert x_1}\otimes \rho^{(2)}_{a_2\vert x_2}\right\}$ admits the 2-LHS  model in Eq.~\eqref{2lhs}. Obviously, the steering boundary $\beta=1$ is attainable from the assemblage $\left\{\rho^{(1)}_{a_1\vert x_1}\otimes \rho^{(2)}_{a_2\vert x_2}\right\}$ with $\eta=\sqrt{3}/3$.

For the parameter range $\sqrt{3}/3\leq \eta \leq\sqrt{2}/2$, the expectations of the state $\rho$ in Eq.~\eqref{isotopic} does not violate the standard CHSH inequality. However, with the LSI, it can be shown that the state $\rho\otimes\rho$ is $2\rightarrow 1$ steerable.

Let us return to the question mentioned at the beginning of this section: The $2\rightarrow 1$ steering can be verified even though Bob performs the SBM which consists of the four rank-one projective operators, $\hat{\Psi}^{\pm}$ and $\hat{\Phi}^{\pm}$, defined in Eq.~\eqref{BSM}. The reason is quite simple: For the three operators $\hat{B}_j$ ($j=1, 2, 3$) in Eq.~\eqref{threepauli}, it can easily be verified that
\begin{eqnarray}
\hat{B}_1&=&\hat{\Psi}^+-\hat{\Psi}^-+\hat{\Phi}^+-\hat{\Phi}^-,\\
\hat{B}_2&=&\hat{\Psi}^--\hat{\Psi}^++\hat{\Phi}^+-\hat{\Phi}^-,\\
\hat{B}_3&=&\hat{\Psi}^++\hat{\Psi}^--\hat{\Phi}^+-\hat{\Phi}^-.
\end{eqnarray}
Instead of the local measurements in Eq.~\eqref{threepauli}, Bob can perform the SBM to detect the $2\rightarrow 1$ steering.

\section{Conclusions}
\label{Sec7}
In this work, we have considered the star network scenario where the central party is trusted while all the edge parties are untrusted. Network steering is defined with an $n$-LHS model. As it has been shown, this $n$-LHS model can be viewed as a special kind of $n$-LHV model. Two different types of sufficient criteria,  nonlinear steering inequality and  linear steering inequality,  have been constructed to verify the quantum steering in a star network. Based on the linear steering inequality, we discussed the case where  the network steering can be demonstrated with a fixed measurement.

The nonlinear inequality in Eq.~\eqref{Bell-like}, which was designed for two-qubit system, works under the constraint that mutually unbiased measurements are performed on Bob's site~\cite{Can3}. In the later work~\cite{Girdhar}, it was proven that this constraint is not necessary in Eq.~\eqref{Bell-like}. In this work, a nonlinear inequality in Eq.~\eqref{non-linear} has been constructed to verify the star network steering. Our inequality can be viewed as a generalization of the one in Eq.~\eqref{Bell-like}. The requirement of mutually unbiased measurements in Eq.~\eqref{BobB} is used in the derivation of Eq.~\eqref{non-linear}. There is an unanswered question here: Can this  nonlinear steering inequality be arrived at without the requirement of mutually unbiased measurements?

As a known fact, a fundamental property is that steering is inherently asymmetric with respect to the observers~\cite{bowles,Midgley}, which is quite different from the quantum nonlocality and entanglement. Actually, there are entangled states which are one-way steerable~\cite{bowles,Bow}. In this work, the steering in the star network is limited to the scenario where only the central party is trusted. Therefore, there are still many unsolved problems, such as how to define network steering in the scenario where the edge parties are trusted while the central party is untrusted.  It is expected that the problems mentioned above can be solved in our future works.

\acknowledgements
This work was supported by the National Natural Science Foundation of China (Grant No.~12147208), and the Fundamental Research Funds for the Central Universities (Grant No. 2682021ZTPY050).

\appendix
\section{Derivation of Eq.~\eqref{non-linear}}
Using the results in Eq.~\eqref{exp21} and Eq.~\eqref{exp22}, one can obtain
\begin{eqnarray}
\label{exp31}
\left\langle\hat{A}_{\bar{0}}\bigotimes\hat{B}_{\bar{0}}\right\rangle&=&\prod_{\mu=1}^n\cos^2\omega_{\mu}\prod_{\nu=1}^n\left(\bm{n}^{(\nu)}_1,T_{\nu}^{\mathrm{T}}\bm{s}^{(\nu)}\right),\\
\label{exp32}
\left\langle\hat{A}_{\bar{0}}\bigotimes\hat{B}_{\bar{1}}\right\rangle&=&\prod_{\mu=1}^n\cos^2\omega_{\mu}\prod_{\nu=1}^n\left(\bm{n}^{(\nu)}_2,T_{\nu}^{\mathrm{T}}\bm{s}^{(\nu)}\right),\\
\label{exp33}
\left\langle\hat{A}_{\bar{1}}\bigotimes\hat{B}_{\bar{0}}\right\rangle&=&\prod_{\mu=1}^n\sin^2\omega_{\mu}\prod_{\nu=1}^n\left(\bm{n}^{(\nu)}_1,T_{\nu}^{\mathrm{T}}\bm{t}^{(\nu)}\right),\\
\label{exp34}
\left\langle\hat{A}_{\bar{1}}\bigotimes\hat{B}_{\bar{1}}\right\rangle&=&\prod_{\mu=1}^n\sin^2\omega_{\mu}
\prod_{\nu=1}^n\left(\bm{n}^{(\nu)}_2,T_{\nu}^{\mathrm{T}}\bm{t}^{(\nu)}\right).
\end{eqnarray}
For $n$ positive parameters $k_{\mu}$ ($k_{\mu}\geq 0$), there exists an inequality
\begin{equation}
\label{math-inequality}
\left(\prod_{\mu=1}^{n}k_{\mu}\right)^{\frac{1}{n}}\leq \frac{1}{n}\sum_{\mu=1}^n k_{\mu}.
\end{equation}
As an application of it, the following result can be obtained
\begin{eqnarray}
&&\left(\left\langle\hat{A}_{\bar{0}}\bigotimes\hat{B}_{\bar{0}}\right\rangle^2\right)^{\frac{1}{n}}
+\left(\left\langle\hat{A}_{\bar{0}}\bigotimes\hat{B}_{\bar{1}}\right\rangle^2\right)^{\frac{1}{n}}\nonumber\\
\label{back} &&\leq
\left(\prod_{\mu=1}^n\cos^4\omega_{\mu}\right)^{\frac{1}{n}}\frac{1}{n}\sum_{\nu=1}^n\left[\sum_{j=1}^2\left(\bm{n}^{(\nu)}_j,T_{\nu}^{\mathrm{T}}\bm{s}^{(\nu)}\right)^2\right].
\end{eqnarray}
For a set of orthogonal unit vectors $\left\{\bm{n}^{(\nu)}_j\right\}_{j=1}^3$,
\begin{equation}
\left(\bm{n}^{(\nu)}_i,\bm{n}^{(\nu)}_j\right)=\delta_{ij},\ \ i,j=1,2,3,
\end{equation}
and the Euclidean norm $\left\vert\left\vert T_{\nu}^{\mathrm{T}}\bm{s}^{(\nu)}\right\vert\right\vert$ can be expressed as $\left\vert\left\vert T_{\nu}^{\mathrm{T}}\bm{s}^{(\nu)}\right\vert\right\vert=\sqrt{\sum_{j=1}^3
	\left(\bm{n}^{(\nu)}_j, T_{\nu}^{\mathrm{T}}\bm{s}^{(\nu)}\right)^2}$. With the fact $\left\vert\left\vert T_{\nu}^{\mathrm{T}}\bm{s}^{(\nu)}\right\vert\right\vert\leq 1$, one can have $\sum_{j=1}^2\left(\bm{n}^{(\nu)}_j,T_{\nu}^{\mathrm{T}}\bm{s}^{(\nu)}\right)^2\leq 1$. By putting it back to Eq.~\eqref{back}, one can obtain
\begin{equation}
\label{exp51}
\sqrt{\left(\left\langle\hat{A}_{\bar{0}}\bigotimes\hat{B}_{\bar{0}}\right\rangle^2\right)^{\frac{1}{n}}+\left(\left\langle\hat{A}_{\bar{0}}\bigotimes\hat{B}_{\bar{1}}\right\rangle^2\right)^{\frac{1}{n}}}\leq\left(\prod_{\mu=1}^n\cos^2\omega_{\mu}\right)^{\frac{1}{n}}.
\end{equation}
Similarly, another inequality can be obtained
\begin{equation}
\label{exp52}
\sqrt{\left(\left\langle\hat{A}_{\bar{1}}\bigotimes\hat{B}_{\bar{0}}\right\rangle^2\right)^{\frac{1}{n}}
	+\left(\left\langle\hat{A}_{\bar{1}}\bigotimes\hat{B}_{\bar{1}}\right\rangle^2\right)^{\frac{1}{n}}}\leq
\left(\prod_{\mu=1}^n\sin^2\omega_{\mu}\right)^{\frac{1}{n}}.
\end{equation}
Finally, using the inequality in Eq.~\eqref{math-inequality} again, the nonliner inequality in Eq.~\eqref{non-linear} can be obtained.

\end{document}